# Destiny: A Candidate Architecture for the Joint Dark Energy Mission

Dominic J. Benford[†1] and Tod R. Lauer[2] for the Destiny Team

1. NASA / Goddard Space Flight Center, Observational Cosmology Laboratory, Greenbelt, MD 20771 USA

2. NOAO, P.O. Box 26732, Tucson, AZ  85726

## ABSTRACT

Destiny is a simple, direct, low cost mission to determine the properties of dark energy by obtaining a cosmologically deep supernova (SN) type Ia Hubble diagram. Its science instrument is a 1.65m space telescope, featuring a grism-fed near-infrared (NIR) (0.85-1.7µm) survey camera/spectrometer with a 0.12 square degree field of view (FOV) covered by a mosaic of 16 2k×2k HgCdTe arrays. For maximum operational simplicity and instrument stability, Destiny will be deployed into a halo-orbit about the Second Sun-Earth Lagrange Point. During its two-year primary mission, Destiny will detect, observe, and characterize ~3000 SN Ia events over the redshift interval $0.4 < z < 1.7$ within a 3 square degree survey area. In conjunction with ongoing ground-based SN Ia surveys for $z < 0.8$, Destiny mission data will be used to construct a high-precision Hubble diagram and thereby constrain the dark energy equation of state. The total range of redshift is sufficient to explore the expansion history of the Universe from an early time, when it was strongly matter-dominated, to the present when dark energy dominates. The grism-images will provide a spectral resolution of $R \equiv \lambda/\Delta\lambda = 75$ spectrophotometry that will simultaneously provide broad-band photometry, redshifts, and SN classification, as well as time-resolved diagnostic data, which is valuable for investigating additional SN luminosity diagnostics. Destiny will be used in its third year as a high resolution, wide-field imager to conduct a multicolor NIR weak lensing (WL) survey covering 1000 square degrees. The large-scale mass power spectrum derived from weak lensing distortions of field galaxies as a function of redshift will provide independent and complementary constraints on the dark energy equation of state. The combination of SN and WL is much more powerful than either technique on its own. Used together, these surveys will have more than an order of magnitude greater sensitivity (by the Dark Energy Task Force's (DETF) figure of merit) than will be provided by ongoing ground-based projects. The dark energy parameters, $w_0$ and $w_a$, will be measured to a precision of 0.05 and 0.2 respectively.

## INTRODUCTION

The discovery of the accelerating expansion of the Universe[1,2] is the most important advance in cosmology since discovery of the cosmic microwave background. The favored explanation of accelerating expansion is that some form of "dark energy" (DE) with negative pressure dominates the mass density of the Universe – yet dark energy itself is unexplained by modern physics. Observational investigations seek to characterize dark energy through precision measurement of the expansion history of the Universe or through tracing the growth of structure. The most precise and accurate techniques can only be accomplished with space-based telescopes.

The SN Type Ia Hubble diagram is the most mature dark energy diagnostic. SN observations led to the original dark energy detection and were promptly employed to provide the first constraints on its equation of state[3]. There are several SN surveys now in progress, and several are planned using ground-based facilities under development. They will provide rich sampling of the low-$z$ Hubble diagram by the time Destiny flies. The ground-based studies will also refine SN as standard candles, improving precision and assessing the systematic effects of chemical evolution and the age of the parent population. However, discovering SN at the highest redshifts needed to constrain DE can *only* be done from space due to the high NIR background on the ground, and obtaining a rich sample of thousands of such events cannot be done with the Hubble Space Telescope (HST) due to its limited FOV.

† Dominic Benford: ph: (1) 301 286 8771; Fax: (1) 301 286 1617; e-mail: dominic.benford@ nasa.gov

Destiny will measure the dark energy equation of state parameter $w(a) = P(a)/\rho(a)$, where $P$ and $\rho$ are the pressure and energy density of the dark energy, and where $a = (1+z)^{-1}$ is the cosmological scale factor. If $w(a)$ varies monotonically, it may be parameterized as $w(a) = w_0 + (1-a)w_a$. The expansion of the Universe accelerates when $w < -1/3$. In a flat universe, with $w(a) = -1$ (dark energy appears as a cosmological constant), the present values of $\Omega_\Lambda \sim 2/3$ and $\Omega_m \sim 1/3$ (the relative contributions of DE and matter to a flat universe, respectively), imply that acceleration is a relatively recent effect, occurring at $z < 1$ when DE began to dominate the Universe. Given this, the expansion should have been slowing down earlier when matter dominated. In fact, this transition has been seen by the Probing Acceleration Now with Supernovae (PANS) program, which used HST to discover eight SN Ia with $z > 1$. Light curves were obtained using Advanced Camera for Surveys (ACS) images, and spectra using the ACS grism. The results[4] show the distinctive change from deceleration at redshift $z > 1$, to acceleration at $z \sim 0.5$ (Figure 1). This is crucial: it demonstrates a change in the sign of the observed effect. Cosmic dust or monotonic changes in the properties of SN Ia with time or chemical composition cannot cause this change in sign. But this is just a start – the precise value of $w$ at any time, or how it may have changed over the history of the Universe, is unknown.

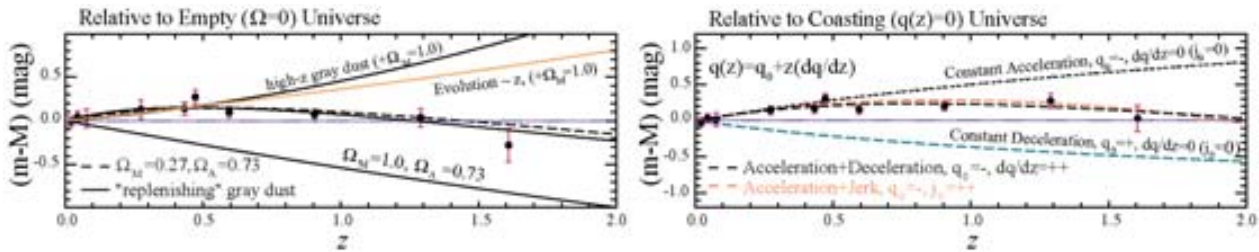

Figure 1 Residuals about the SN Ia Hubble diagram assuming an empty Universe (left panel) or a coasting Universe with no acceleration or deceleration (right panel). The points at z ~0.5 are too dim, showing the effect of accelerating expansion, while the points at z ~1.5 are bright, showing that this epoch (when matter dominated) was one of deceleration. Simple evolution or dust extinction cannot explain this transition. Figure after Riess et al. (2004)[4].

Destiny will provide the data needed to extend the more precise near-IR Hubble diagram into the redshift range at the onset of acceleration, where information on the cosmic expansion history has the largest leverage on dark energy constraints. Destiny will obtain a rich sample of SN Ia at $0.4 < z < 0.8$ that will cross-calibrate and enhance the SN sample from low-redshift ground SN surveys, which will be used in the analysis for $0 < z < 0.4$. For supernovae at $z > 0.8$, observations from the ground become difficult for optical detectors and are beyond current technology for the infrared (IR). Destiny will provide the unique information at $z > 0.8$ that is critical for showing the transition from a matter to dark energy dominated universe. The $z = 1.7$ redshift end of Destiny's sample serves as an "anchor" for determining variation in $w$ during the epoch when acceleration occurs[5], and provides an important constraint on more complex behavior of $w_a$. The high redshift observations also provide the most stringent tests for evolution of the SN Ia properties with cosmic epoch[6].

## 1. THE DESTINY SUPERNOVA SURVEY

Over a nominal two year mission, Destiny will observe SN by repeatedly imaging an unbiased field near each ecliptic pole. Using two separate fields partially mitigates the cosmic variance contribution from large scale structure. The SN spectra will be isolated by image differencing. This technique sidesteps all issues of overlap of the SN Ia spectra with that of their host galaxies, and any other concerns about source confusion. The full octave of Destiny's wavelength coverage is sufficient to correct the SN photometry for extinction, and its time coverage is sufficient to measure the light curve decay-rate corrections.

For each three month interval, the spacecraft roll angle will be held fixed as the Earth and the spacecraft at L2 orbits the Sun. It will then roll 90° about the ecliptic poles for the next three-month interval (as only one side of the spacecraft can face the Sun). Over each interval, images will be obtained at a steady tempo to track the rise and fall of SN over their light curves. The survey fields are subdivided into integral Destiny FOV locations. The repeated visits (18 per roll) will be done with identical pointing for each FOV so the SN spectra, host galaxy light, and other objects always fall on the same pixels. This allows accurate subtraction of non-varying sources by difference imaging without flat fielding.

Precise flats can then be constructed for the isolated SN spectra, once their wavelength calibrations have been measured. The reference image for each roll comes from stacking the 18 images obtained at each FOV location during the opposite year at the same roll angle. However, while the images are associated by roll angle, all SN will be observed continuously, regardless of how the roll may change, once the SN spectra have been isolated.

Optimal extraction of SN spectra from the first year will begin after the first visits at all rolls are complete. However, brighter SN will always be visible without image differencing, and during the first year, differencing imaging among subsets of the data will be done to provide immediate identification and evaluation of the SN events. Events in the second year will be optimally extracted as they occur.

Because Destiny focuses on SN Ia at $z > 0.4$, the $(1+z)$ time dilation eases the tempo of repeat visits needed to track the luminosity evolution of the SN compared to a mission that also observes SN Ia at low redshifts. Monte Carlo simulations show that a five-day sampling interval is sufficient. The total exposure time is 4 hours for each 0.12 square degree Destiny field, which will be divided into several sub-exposures to allow for cosmic-ray events, as well as to perform sub-pixel dithering to generate well-sampled images. Operating Destiny at L2 allows continuous observation of fields near the ecliptic poles. The on-target time fraction should approach unity, allowing 1.6 square degrees to be imaged in ~2.5 days near each ecliptic pole. Observation of both fields will be alternated within a 5.1-day block. The total survey area is 3.2 square degrees, composed of 28 separate fields. A small portion of each block will be used for direct J-band imaging of the fields to establish the zero-point of the wavelength scale of the SN spectra, as well as the morphology of and location of the SN in their host galaxies.

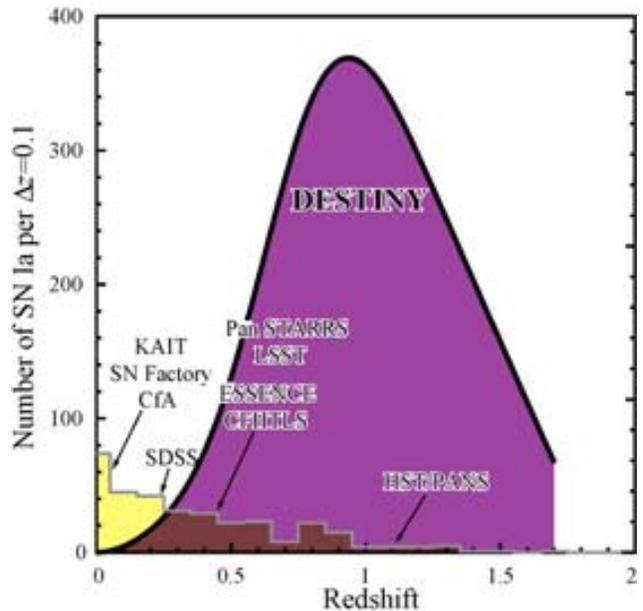

Adopting SN Ia rates from Dahlen et al. (2004)[7], for $\Omega_0 = 1$, $\Omega_\Lambda = 0.7$, and $H_0 = 72$ km/s/Mpc, implies that >3000 SN Ia within $z < 1.7$ should be observed within a two year period. At $z = 0.5$, ~100 SN Ia are expected in a $\Delta z = 0.1$ bin. The yield increases to ~400 events at $z = 1.0$, declining to ~100 events at $z = 1.5$ (Figure 2). This sample yields a Hubble diagram accurate to ~1% on the $\Delta z = 0.1$ scale, given the 0.14 magnitude dispersion in SN Ia peak luminosity that remains after the light curve decay-rate correction has been applied. There should be at least 100 SN per $\Delta z = 0.1$ bin for $z < 0.5$ by the time Destiny launches; it will build on this low redshift SN Ia sample.

Figure 2. Destiny will observe > 3000 SN Ia at $z < 1.7$ in two years. The black histogram shows existing SN Ia detections with redshift. Future and on-going SN surveys are listed by the approximate redshift range that they sample.

## 2. THE DESTINY WEAK LENSING SURVEY

After concluding the Destiny SN Ia survey, the spacecraft will conduct a weak lensing survey by using Destiny's instrument in its direct-imaging mode. Weak lensing provides a complementary and powerful probe of dark energy. Shapes of galaxies are slightly distorted as the light travels through the density fluctuations along the line-of-sight to the observer. The statistical properties of galaxy shapes thus depend on the statistical properties of these density inhomogeneities. Dark energy constraints from WL are primarily due to a Cosmic Microwave Background (CMB)-calibrated standard ruler in the power spectrum of the density inhomogeneities. By studying weak lensing as a function of the redshifts of the source galaxies, it is possible to reconstruct the angular diameter distance as a function of redshift. In addition, WL is sensitive to the growth of structure, and thus is essential for discriminating between dark energy and modified gravity as explanations for the acceleration.

The Destiny WL survey strategy is to concentrate on measurements for which being in space presents a strong advantage. As a diffraction-limited space telescope, Destiny will have a small, stable point spread function (PSF) and the ability to deeply image the sky in NIR bands. The small PSF increases the number density of galaxies whose shapes can be measured, reducing statistical noise. The stability of the PSF controls systematic errors in the shape measurements. Shape measurements in the J band and photometry in the z, J, and H bands will be done from space, while photometry in optical bands, necessary for the photometric redshift determinations, can be done from the ground with facilities such as the Large Synoptic Survey Telescope (LSST). The NIR/optical combination greatly improves the accuracy of the photometrically-determined redshifts, particularly in the $1.4 < z < 2.5$ "redshift desert" in which neither the Balmer nor Lyman breaks are in the window of ground-accessible optical bands. As with the SN survey, measurements at $z \sim 1.5$ provide important leverage on the overall form of $w(a)$. Despite these advantages, the Destiny team acknowledges that the merits of ground-based versus space-based WL studies are still being debated in the community. The use of Destiny for WL surveys does not interfere with its primary SN survey, and instead is a "bonus" capability.

A one-year WL survey will cover 1000 square degrees in three IR bands, achieving a (5σ) limiting J-band magnitude of 26. The useful source density at this limit is 180 galaxies per sq. arcmin., thus the survey will contain over 600 million galaxies. The final accuracy of the WL survey will depend on the control of systematic errors and the quality of theoretical predictions. Important sources of systematic error are the effects of intrinsic galaxy shapes and alignments.

A Destiny WL survey will provide an unprecedented sample of cluster galaxies at moderate red- shift with nearly HST-quality photometric information about stellar content and star formation, with no extra cost or operations requirements. Cluster counting and the assessment of baryonic and mass/light fraction in clusters of galaxies provide useful constraints on the dark matter and dark energy contents of the universe, which are independent of and complementary to constraints from SN and WL studies. Destiny will detect over-densities, traced by the light of galaxies, as a function of galaxy type and color. The science team will investigate how a 3D galaxy clustering analysis of the Destiny WL survey data can provide additional constraints on dark energy.

## 3. CONSTRAINING THE DARK ENERGY EQUATION OF STATE

The predicted constraints on the dark energy equation of state parameters $w_0$ and $w_a$ to be provided by the Destiny mission (with SN measurements tied to ground-based observations) are shown in Figure 3 (courtesy of Lloyd Knox and Yong-Seon Song). This analysis assumes that the ground-based surveys will provide ~100 SN Ia per $\Delta z = 0.1$ bin for $z < 0.7$ prior to the Destiny mission, that CMB data will be available from Planck, $H_0 = 72 \pm 8$ km/sec/Mpc, and that the Universe has a flat spatial geometry. In addition, it is assumed that SNe are standard candles with a 0.14 statistical apparent magnitude RMS dispersion. A drift of 0.02 magnitudes in the absolute flux scale over $0 < z < 1$ is allowed for as a plausible systematic error.

Contours for a 1000 square degree Destiny WL survey, and the combination of this survey with Destiny's baseline SN surveys, are also shown in Figure 3. The high degree of complementarity between the SN and WL techniques is evident, with the combination achieving a much tighter constraint than is obtained by either technique individually. The combination also provides robustness to some modeling assumptions: the area of the $w_0$, $w_a$ SN-only contour increases significantly if the assumption of a flat Universe is abandoned, but by less than 5% for the combined SN and WL case (see right panel of Figure 3). The analysis of the space-based weak lensing survey follows the procedures presented by Song & Knox (2004)[8]. A larger survey area decreases the sample variance of the WL power spectra and greater depth improves the photometry for redshift determination and increases the numbers of galaxies that can be used, thereby reducing the shot noise in the inferred shape power spectra. The analysis is conservative, using only shear two-point function statistics. Tighter constraints may be possible using higher-order correlations[9], shear-galaxy correlations[10], and shear peak statistics[11].

The combined Destiny SN and WL surveys will achieve an order-of-magnitude reduction in the area enclosed by the $w_0$, $w_a$ contour shown in Figure 3, as compared to enclosed area expected after completion of ongoing projects, such as Canada-France-Hawaii Telescope Legacy Survey (CFHTLS) and Equation of State: SupErNovae trace Cosmic Expansion (ESSENCE). Destiny's results will be an exciting advance in our capability to detect non-cosmological constant dark energy or to otherwise distinguish between theories.

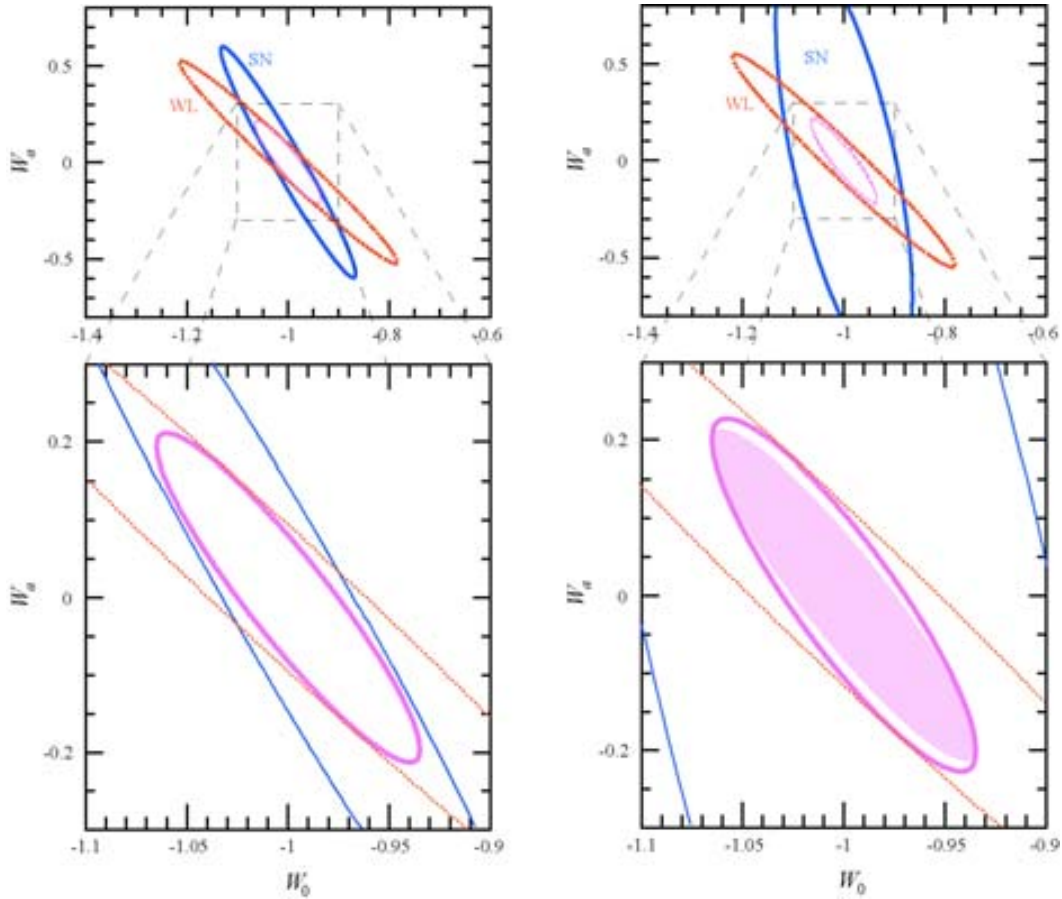

Figure 3. (Left) The SN and WL surveys obtain separate 1σ limits on the DE $w_0$ and $w_a$ parameters, with the assumption of a flat Universe (upper panel). The combination of the two surveys (lower panel) gives especially powerful constraints for $w_0$ and $w_a$. (Right) Constraints on $w_0$ and $w_a$ (upper panel), now allowing for curvature. The joint constraints (lower panel) are only slightly affected by curvature; the flat-Universe model is shown as a shaded oval. Even though the area of the SN survey contour greatly increases with curvature, it still works with the WL survey to reduce the total area of the joint $w_0$ and $w_a$ contours.

## 4. SUPERNOVA SPECTROPHOTOMETRIC OBSERVATIONS

The Destiny approach, using grism imagery, yields simultaneous SN photometric measurements and spectra over the entire Destiny bandpass. This strategy can be contrasted with the more conventional approach of observing the SN through a suite of broadband filters. Although a typical single $R=5$ filter has a lower background than the grism, simulations show that when the necessary number of filters that must be scanned through in succession is eight or more, the grism reaches a given signal-to-noise (S/N) ratio more quickly for most SN events and with greater operational simplicity than a space telescope of equivalent size using conventional broadband imaging. The HST/ACS grism is already in use to obtain spectra of high-redshift SN[4], as well as to discover them initially[12].

The grism's greatest strength is its ease and simplicity in capturing SN spectra at the time of maximum brightness. Spectra are needed to classify the SNe into subtypes and to measure their redshifts. If the SN were observed one at time with a conventional single-object spectrograph, this operation would actually require a majority of the spacecraft time. With Destiny, the spectra are not only sampled at maximum, but at all phases of the SN light curve.

The Destiny spectral resolution (80Å/pixel) is more than sufficient to distinguish the identifying characteristics of the various SN types. As a result of large expansion velocities, the features in SN spectra are quite broad. Type II SN are characterized by hydrogen lines, while Type I SN lack hydrogen and are subdivided into three classes. The key identifying feature of SN Ia, the 6150Å silicon line, will be apparent in Destiny spectra for $0.4 < z < 1.7$, setting the nominal range of its SN survey. The absorption lines of helium appear in SN Type Ib. Type Ic SN lack all of these features. The ability to distinguish the SN subtype is critical for cosmological uses, because only Type Ia SN are sufficiently well understood to be used as reliable standard candles. Fortunately, the spectra near maximum are clearly distinctive among all SN types. The tempo of Destiny observations will result in well-sampled light curves, ensuring that all SN will have spectra at maxima, yet without requiring time-critical interactions with the spacecraft. The spectra will also help to explain the intrinsic dispersion in Type Ia properties. Within the Type Ia class, some variation in luminosity is reflected in spectroscopic differences. One of the indicators of a subluminous Type Ia is the relative strength of an absorption line near 5800Å. This feature will be measured in 95% of Destiny spectra for the expected redshift range. Thus, the spectra help to distinguish unusual Type Ia SN in the sample.

Preliminary simulations have established Destiny's ability to obtain photometric measurements of the needed precision. Figure 4 (left image) shows a simulated deep galaxy field with realistic galaxy clustering, spectral energy, luminosity, and redshift distributions, observed with the Destiny spatial resolution and bandpass. Sizes and morphologies were drawn from cutouts of real galaxies detected in deep HST/ACS imagery. Figure 4 (middle image) shows the same field as observed by Destiny with the 4-hour single-epoch exposure. The dominant noise source is the sky background, which is the integral of the zodiacal light over the full spectral bandpass (provided by Giavalisco et al. 2002[13]). Figure 4 (right image) shows the simulated detection by image differencing of a z=1.5 SN Ia at peak luminosity, where the input spectrum is provided by the models of Nugent et al. (2002)[14].

Figure 5 (left) shows the simulated S/N per each 80Å spectral pixel in a 16-hour exposure (four epochs) for SN Ia near maximum at various redshifts. R~75 resolution is clearly adequate to characterize the broad spectral features. At $z > 1.4$, the S/N per pixel becomes low at the redder wavelengths. Destiny will be helped there by two factors: one, time dilation will allow spectra from multiple visits to be co-added, while still maintaining the same time resolution as $z$ ~0.5 events; and two, S/N can be increased by sacrificing some time resolution at high redshifts. Accurate photometry can be obtained at all redshifts over the time of maximum light, and deep enough into the decaying portion of the light curve to characterize the decay rate. Figure 5 (right) summarizes the total photometric S/N integrated over the entire Destiny bandpass per 4-hour visit for the SN Ia, as a function of redshift and time since maximum light. The simulations included the full effect of observing the SN against their host galaxies, as well as the overlapping spectra of other galaxies in the field. The zodiacal background is by far the strongest source of noise; on average, the host and background galaxies increases the noise by only 8%.

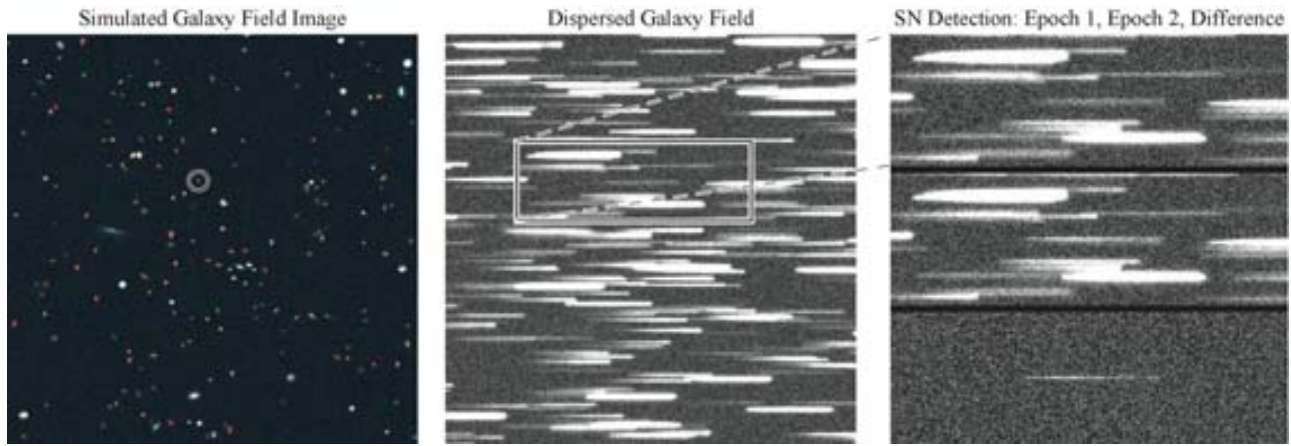

Figure 4. Grism simulations start with input field galaxy samples with realistic luminosity, redshift, and spectral energy distributions. At left is a simulated galaxy field covering 512×512 pixels (1.3 arcmin×1.3 arcmin), only 0.4% of the total instantaneous field of view; one galaxy (circled) is a SN host galaxy. In the center is a dispersed grism image for a 4-hour exposure of the field, where blue is dispersed to the left. The method of isolating SN spectra by image differencing is shown at right: the upper panel is the central portion of the simulated field shown previously; the middle panel is a second, similar grism image with a $z$=1.5 SN Ia included in one host galaxy; the bottom panel shows the SN isolated after subtracting the first from the second.

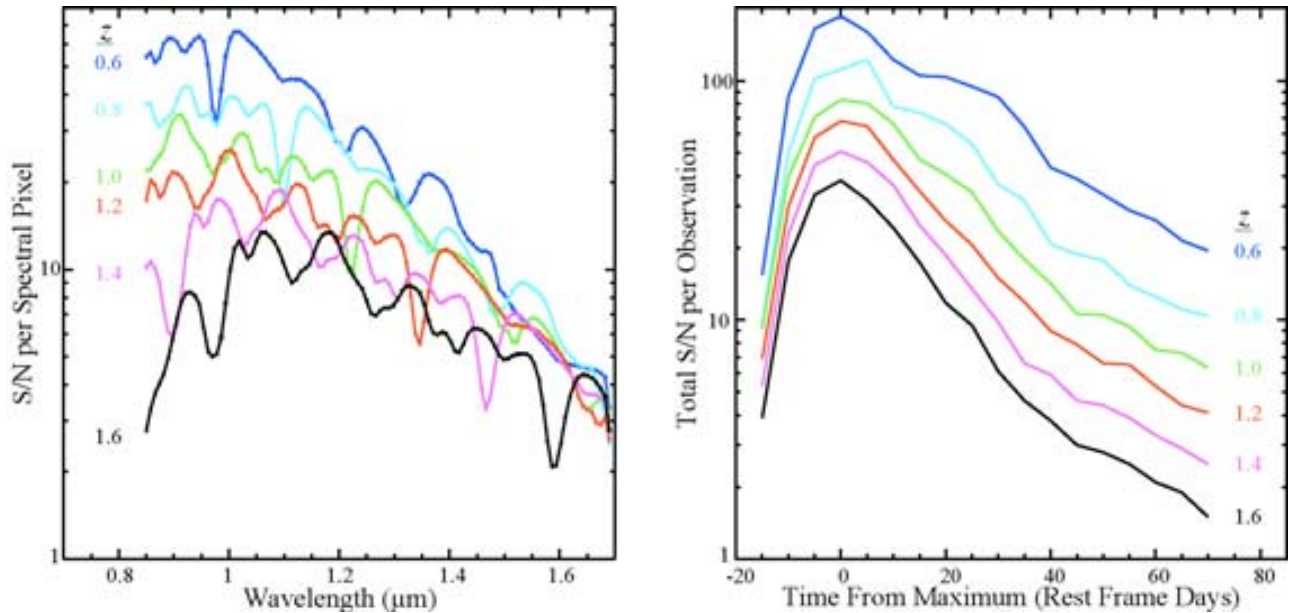

Figure 5. (Left) Destiny SN spectra at maximum light have sufficient S/N and resolution to derive redshifts and type classification. S/N per 80Å spectral pixel is shown with redshift and wavelength for spectra coadded from four epochs surrounding maximum light. (Right) The Destiny photometric observations of the SN will have sufficient signal/noise to observe the decay of the SN events well past maximum light. Signal-to-Noise ratio of the total SN flux detected over the Destiny bandpass as a function of redshift and time since maximum light is shown.

There are multiple methods for using Destiny photometry to generate the SN Ia Hubble diagram. The grism spectrometry offers great flexibility in how the SN photometric measurements are extracted. For example, the photometry can be binned into $R=5$ bands that shift with redshift, completely eliminating the need for K-corrections. Yet, the data also allow the more traditional NIR bands to be adopted, while using the Destiny spectra to derive more precise K-corrections than could be done with broadband photometry alone. The optimal solution may involve a finely weighted optimized integration of the spectra over the entire bandpass, avoiding conversion of the results to standard bands altogether. Best of all, this flexibility gives the best path towards incorporating any new methodologies of SN photometry that may emerge from the ongoing investigations.

## 5. DESTINY MISSION DESIGN

The Destiny requirements definition is based on justifiable scientific measurement goals combined with technical feasibility and a simple, low-cost approach with high heritage. Destiny's approach results in an iterated requirements list that reflects a combination of both the science goals and mission design. The three primary drivers of the Destiny instrument and mission design are: the requirement to obtain spectra of all SN events near maximum light for redshift and type classification; to accomplish this in a way that does not detract from performing continuous photometric monitoring of the SN events; and, to create a mission structure that does not demand continuous time-critical analysis or commanding. Combining these requirements, the Destiny team has selected grism-imaging as the dominant mode of operation. This methodology completely eliminates time-critical operations as a cost driver. It also maximizes "on-survey" time, which reduces the required FOV and primary mirror size. This approach also has the advantage of eliminating the systematic error source from K-corrections associated with broad-band filters. In keeping with its philosophy of minimizing mission complexity, the Destiny team has constrained the requirements to fit within the overall mission scope. After a description of the mission, Table 1 summarizes key parameters of Destiny.

### 5.1 Science instrument

The driving force behind the Destiny design architecture is simplicity of design, low cost, and the use of heritage hardware and processes. The Destiny science payload consists of an optical telescope assembly (OTA), aft optics, electronics and a science focal plane array (FPA), and a guide focal plane (guide camera). The guide camera provides pointing error estimates to the spacecraft bus after initial target acquisition and enables the fine pointing required for the science mode. Thermal control of the science payload is completely passive and utilizes a dedicated thermal radiator. A simplified block diagram of the Destiny instrument is shown in Figure 6.

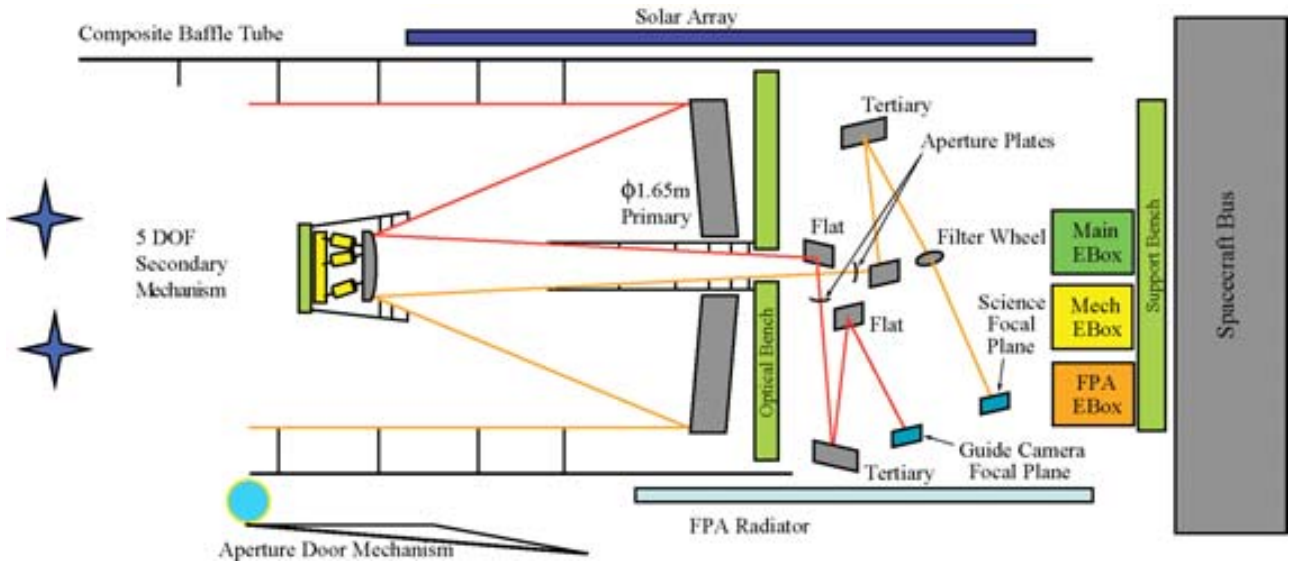

Figure 6. Destiny science instrument block diagram. The Destiny instrument is a simple, straightforward design that utilizes high-heritage components to reduce risk and cost.

### 5.1.1 Optical Telescope Assembly and Aft Optics:

The Destiny optical system (Figure 7) is a three-mirror anastigmat imaging system consisting of a primary mirror ellipsoid, a convex secondary, and two tertiary mirrors. The primary mirror is a lightweighted, sandwiched, and frit-bonded ultra low expansion glass (ULE) mirror, providing strength and mass margin. The remaining optics are also made of ULE, with moderate light-weighting. The primary, secondary, and tertiary mirrors work together to form a well-corrected, flat-field image. However, the telescope forms an intermediate image, which in itself is not well corrected, nor on a flat plane. It is at this intermediate image location that the field stops, or entrance aperture plates, for each channel are placed. The fields are cut from curved surfaces, matching the focal plane of the telescope, and generate well-defined field stops. The tertiary mirrors have the same prescription but differ in size; one is located in the optical path of the science instrument, and the other in the guide camera optical path.

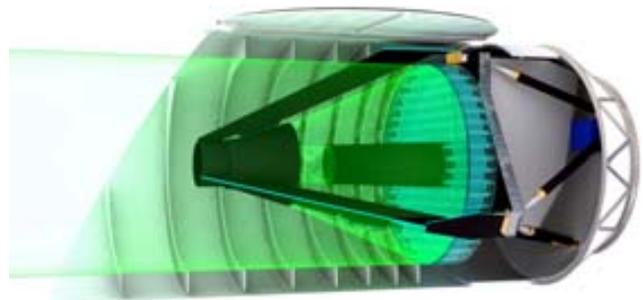

Figure 7. The Destiny OTA is a simple design with high performance, that enables numerous optical corrections via a powered secondary mirror support.

In the science channel of the focal plane (Figure 8), the light beam is folded immediately after the entrance aperture plates and relayed to the tertiary mirror (TM). The TM re-images the telescope image directly to the FPA. Between them is a pupil, where a stray light stop can be placed. This is also the best location for filters, as the field is affected uniformly, and the beam size is minimized as an image of the primary mirror is formed. The filter wheel is placed at this location. The filter wheel holds bandpass filters as well as a grism (diffractive plus refractive) dispersive element. In the guide field, the light is folded prior to the entrance aperture plate and relayed straight through to the tertiary mirror. For packaging, there is an additional fold mirror directly after the pupil, allowing the guide camera FPA to be located near to the science FPA. The placement of the guide detectors is symmetric to the science focal plane (see Figure 8). The guide field and the science focal plane will then have equivalent optical positions. This will allow the same tertiary mirror prescription to be used for both focal planes. The guide optical path is coplanar with the science optical path and in-band tracking is utilized so that a beamsplitter is not required. The image is close to diffraction limited at 850nm across the full science field.

The primary mirror assembly mounts provide the structural interface between the OTA and the optical bench, and provide a load path to the structural ring. The mounting design employs a bipod flexure system that provides a stress-free mounting for the mirror. The bipod mounts will be optimized to minimize gravity vector distortion, and tailored to the natural frequency of the optic. There is extensive industry experience using the bipod flexure system, and systems of this type have been built and successfully flown numerous times. The secondary mirror is mounted in a ring supported by a metering structure (see Figure 7) and is powered by a five degree of freedom mechanism. The mechanism allows for tip, tilt, defocus and de-center to correct misalignments and de-focus caused by thermal distortions or mechanical shifts induced at launch. A barrel, which operates as a sunshade, surrounds the mirrors and metering structure to block light and provide thermal protection. An aperture door closes off the open end of the sunshade to protect against contamination during ground testing, launch and transit to L2.

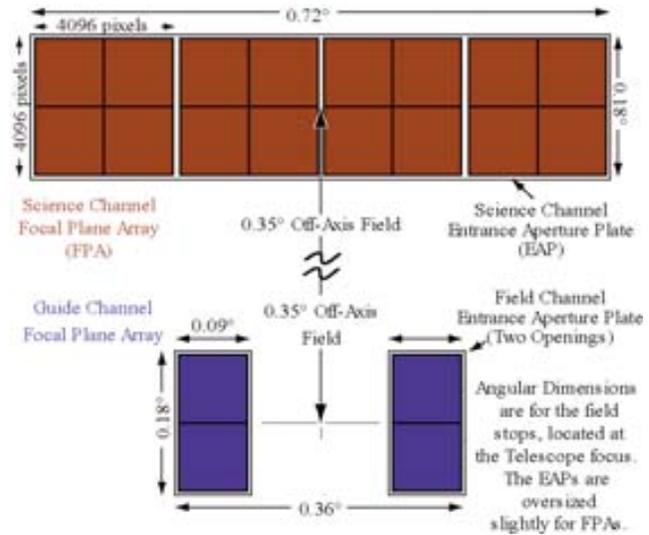

Figure 8. The focal plane layout provides high quality wide field imaging for both the science and guide FPAs and permits the use of identical Tertiary mirror.

### 5.1.2 Electronics:

The instrument electronics distributes power within the instrument and provides for control of all instrument functions. The main electronics box provides general control functions, power functions and interfaces with the spacecraft bus. The focal plane electronics control setup and readout of the focal planes and oversee transfer of the data to the spacecraft. The electronics do not process the science data; lossless compression is performed on the spacecraft bus by the solid state recorder. The mechanism drive electronics control the functions of and provide conditioned power to the three mechanisms used in the instrument: the secondary mirror mechanism, the aperture door, and the filter wheel.

### 5.1.3 Focal Plane Array:

The science focal plane array (Figure 8) utilizes 2k×2k hybridized HgCdTe detectors arranged in a rectangular 4k×16k focal plane array (64Mpixels). This array provides a FOV of 0.175 by 0.712 degrees. Candidate detectors for the Destiny mission are currently being developed for James Webb Space Telescope (JWST) (by Rockwell) and for Variability of solar IRradiance and Gravity Oscillations (VIRGO) (by Raytheon), and would require little or no new development for use by the Destiny project. The detectors are maintained at 140K, a temperature sufficient to reduce thermal noise, and which can be obtained easily using passive cooling methods. The most significant area of risk in the detector implementation is packaging and layout of the focal plane. To minimize this risk and ensure a complete understanding of the risk areas, the Destiny concept study will fund a development model of the focal plane array to optimize packaging strategies and ensure low risk for the Destiny flight implementation.

### 5.1.4 Guide Camera:

The guide camera generates extremely accurate line of sight (LOS) angles to sense sub-pixel pointing and motion during science data acquisition. The angles are sent as quaternions to the spacecraft control system. Locating the guide camera detectors symmetrically to the science focal plane (Figure 8) allows the guide camera to detect both platform motion and intra-OTA component motions. This enables the guide camera to make a very accurate and direct measurement of the OTA's line of sight. The guide camera parameters are shown in Table 1.

Table 1. Key parameters describing the Destiny instrument and mission.

| Category | Parameter | Value |
|---|---|---|
| Requirement | SN Detection | S/N >7 at 20 days past maximum |
| | SN Redshifts | Accuracy of Δz~0.005, achieved via spectral lines |
| | SN Number / Survey Area | 3000 SNe / 3 sq. deg. (based on rates at $z$~1) |
| | WL Survey Imaging | Filters for broadband imaging, grism |
| Instrument Optical Parameters | Instantaneous FOV | 0.175° × 0.712° |
| | Telescope diameter, design | 1.65 m; three-mirror anastigmat made of ULE glass |
| | Angular resolution | 0.15″ at 1μm (diffraction limited) |
| | Plate scale | 0.15″/pixel |
| | Wavelength range | 0.85-1.70μm |
| | Spectral resolution | R ≡ λ/Δλ = 75 @ 1.2μm |
| | Detector format | 4k × 16k pixels |
| | Detector type | HgCdTe; hybrid 2k × 2k arrays |
| Instrument Resources | Mass | 1784 kg (instrument only; observatory is 2509kg) |
| | Power | 785 W |
| | Volume | 4.4 m long; 2.5 m diameter |
| | Data volume | 37 GB/day |
| | Thermal control | Passive cooling of detectors; active heating of optics |
| Attitude Control | Guide camera FOV | 100 sq. arcmin. |
| | Guide sample rate | 10 Hz |
| | Jitter / Drift | <30 mas / 10 mas total |
| Mission Parameters | Lifetime | >3 years |
| | Launch date | 2013 |
| | Orbit | Sun-Earth L2 halo |
| | Launch vehicle | Delta IV 4040-12 or Atlas V 401 |

**5.2 Spacecraft and Launch**

With the overall goal of a high Technology Readiness Level (TRL), high heritage, low risk mission, the Destiny spacecraft design draws heavily from previous and current missions and hardware, and is a simple, straightforward design using proven technology.

The most stringent requirement for the spacecraft is that pointing must be accurate to 0.010 arcsec (1σ) and stable to better than 0.030 arcsec over the 900sec duration of a single integration. The spacecraft bus also provides the necessary propulsion for momentum dumps and to maintain orbit at L2 and components needed to execute pointing maneuvers, process commands, handle science and housekeeping data communicate with the ground stations, and provide power to the observatory. The spacecraft construction is straightforward, utilizing an aluminum structure with the instrument kinematically mounted to the top deck of the spacecraft. Power is provided by body-mounted solar arrays on five of the eight spacecraft bus sides. The bus utilizes a standard passive thermal control system, consisting of coatings, multilayer insulation, heaters, and thermostats, with thermal isolation between the science instrument and spacecraft bus. The communications system is redundant and uses only heritage components. Ranging, commands, and telemetry are carried by an S-band system, while the data is downlinked using a Ka-band system.

Destiny will operate in a Sun-Earth L2 halo orbit at a distance of ~1.5·10$^6$ km from Earth. It is inertially pointed using star trackers; a hydrazine propulsion system provides station-keeping thrust. The launch mass is 2509 kg (with margin). This mass can be carried to L2 by either a Delta IV 4040-12 or an Atlas V 401 launch vehicle, wither of which can accommodate the 4.4 m-long, 2.5 m-diameter payload.

### 5.3 Operations, Management, and Schedule

The Destiny mission operations concept takes advantage of existing NASA infrastructure and capabilities to communicate with the spacecraft, acquire telemetry and data, and to distribute this information to participating groups and centers. The Destiny Ground System is composed of DSN ground stations, a Mission Operations Center (MOC), a Science Operations Center (SOC), and a Destiny Science Analysis Center (DSAC). The Tracking and Data Relay Satellite System (TDRSS) constellation and the White Sands complex will be used in launch and early operations only. The MOC is responsible for real-time health and safety processing, including the management of safemode events; spacecraft commanding; mission planning and scheduling; instrument data handling and engineering data trending/analysis; orbit determination and control; network and contact scheduling; spacecraft monitoring and control; mission operations testing; and Level 0 data processing and distribution. The SOC is responsible for observer support; science data trending; data validation and calibration, including Level 1-2 data processing of all data (to produce spectrophotometric images); Level 3 processing of the generic (non-prime science) data; and maintenance of a public archive of non-prime science data. The DSAC is responsible for Level 3 processing of the prime science data. This includes identifying and isolating supernova events; extracting and calibrating event spectra; producing a spectra vs. time graph of each individual supernova; and archiving and distributing data to the astronomical community.

Since JDEM is a joint NASA/Department of Energy mission, there is already a notional management structure for the JDEM Dark Energy Investigation and mission. Overall management of Destiny is held at the JDEM Project Office; while the Dark Energy Investigation (which consists of the scientific aspects of the mission and the science instrument), is led by the PI, Dr. Tod R. Lauer of the National Optical Astronomy Observatory (NOAO). The Dark Energy Investigation development is conducted with NASA / GSFC as the lead center, with Dr. Dominic J. Benford as the Institutional PI; other key GSFC members include John Galloway, Ruthan Lewis, and Rud Moe. The Destiny Science Team consists of an additional 12 Co-Investigators and 14 Collaborators from universities, NASA centers, and DOE centers: Matthew Beasley, Kenneth Carpenter, Chris Fassnacht, Jay Holberg, Robert Kirshner, Lloyd Knox, Sangeeta Malhotra, Marc Postman, James Rhoads, Nicholas Suntzeff, Thomas Vestrand, Robert Woodruff, Megan Donahue, Wendy Freedman, Chris Fryer, Aimee Hungerford, Lori Lubin, Tom Matheson, Phillip Pinto, Yong-Seon Song, George Sonneborn, Sumner Starrfield, Frank Timmes, Mike Warren, Rogier Windhorst, and Ann Zabludoff. Six industry partners have roles in the design and development of the Destiny mission.

The preliminary Destiny mission development schedule is based on a project start date of October 2008. A total lifecycle of eight years is divided into five years for mission development and three years for the dark energy phase science operations. Preliminary analysis (Phase A) lasts one year, followed by a one year definition period (Phase B); the design (Phase C) and development (Phase D) require nine months and two years, respectively. After launch a three month period to cruise to L2 and perform check-out is required prior to the three year mission operations (Phase E).

### 6. SUMMARY

The Dark Energy Space Telescope, Destiny, is an all-grism NIR 1.65 m survey camera optimized to return richly sampled Hubble diagrams of Type Ia supernovae (SN) over the redshift range $0.4 < z < 1.7$, for characterizing the nature of the recently-discovered "dark energy" component of the universe. SN will be discovered by repeated imaging of a 3.2 square-degree area located at the ecliptic poles. Grism spectra with resolving power $\lambda/\Delta\lambda = 75$ will provide broad-band spectrophotometry, redshifts, SN classification, as well as valuable time-resolved diagnostic data for understanding the SN explosion physics. This mission methodology features only a single mode of operation with no time critical interactions, a single detector technology, and a single instrument. Although grism spectroscopy is slow compared to SN detection in any single broad-band filter for photometry, or to conventional slit spectra for spectral diagnostics, the multiplex advantage of being able to obtain images and spectra for a large field-of-view simultaneously over a full octave in wavelength makes this approach highly competitive. Additionally, this instrument can perform a wide-area weak lensing survey to add complementary constraints to the characterization of dark energy. Destiny is currently a proposed mission for NASA's "Concept Studies for the Joint Dark Energy Mission" endeavor. If ultimately selected as proposed, it would commence in late 2008, launch in 2013 (Figure 9), and complete its surveys in 2016. Combined, the surveys will have more than an order of magnitude greater sensitivity than will be provided by ongoing projects. The dark energy parameters, $w_0$ and $w_a$, will be measured to a precision of 0.05 and 0.2 respectively.

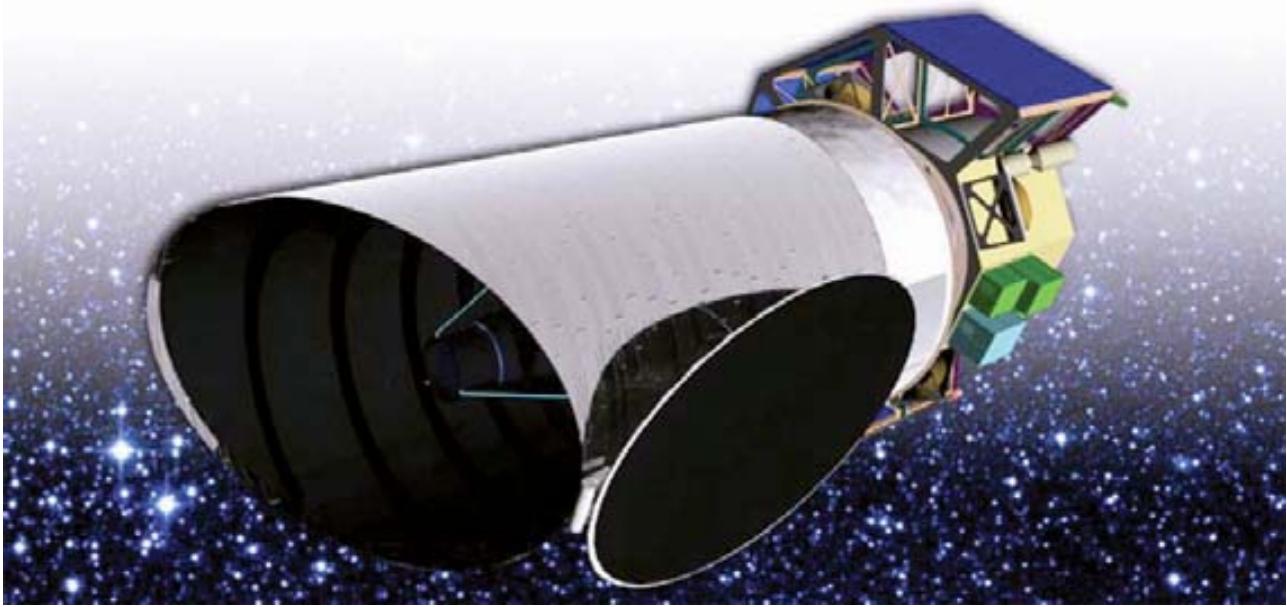

Figure 9. Artist's concept of Destiny in operation at L2 during its three-year survey mission.